\documentclass{article}

\usepackage{PRIMEarxiv}

\usepackage[utf8]{inputenc} 
\usepackage[T1]{fontenc}    
\usepackage{hyperref}       
\usepackage{url}            
\usepackage{booktabs}       
\usepackage{amsmath}
\usepackage{amsthm}
\usepackage{amssymb}
\usepackage{amsfonts}
\usepackage{mathtools}
\usepackage{amsfonts}       
\usepackage{nicefrac}       
\usepackage{microtype}      
\usepackage{fancyhdr}       
\usepackage{graphicx}       
\usepackage{times}
\usepackage{subfig}
\usepackage{xcolor}
\usepackage{array}
\usepackage{cite}
\usepackage{multirow}
\usepackage{adjustbox}
\usepackage{algorithm}
\usepackage{algpseudocode}
\usepackage[font=footnotesize,labelfont=bf]{caption}
\usepackage{tikz}
\usetikzlibrary{arrows.meta,positioning,calc,fit,backgrounds}

\usepackage{enumitem}

\newcommand{\R}{\mathbb{R}}

\newcommand{\norm}[1]{\left\lVert #1\right\rVert}

\newtheorem{theorem}{Theorem}
\newtheorem{lemma}{Lemma}
\newtheorem{proposition}[theorem]{Proposition}
\newtheorem{corollary}{Corollary}

\newtheorem{assumption}{Assumption}

\newtheorem{remark}{Remark}

\title{Deep Reinforcement Learning with Smith-Predictor Delay Compensation
}

\author{
  Moh Kamalul Wafi \\
  Department of Electrical and Computer Engineering \\
  Northeastern University \\
  Boston, MA, USA\\
  \texttt{\{wafi.m\}@northeastern.edu} \\
}

\begin{document}
\maketitle

\begin{abstract}
Time delays can degrade feedback performance and destabilize
a closed loop that is stable without delay. This paper studies
Smith-predictor-assisted reinforcement learning for systems
with constant input delays. A deep deterministic policy
gradient (DDPG) actor is trained on a nominal delay-free
environment and deployed with a Smith predictor using a fixed
nominal delay. For linear controllers, we derive exact
closed-loop maps under plant and delay mismatch and provide
small-gain and $\nu$-gap robustness conditions. A separate
delay-independent Razumikhin condition gives a sufficient
stability test for a fixed Lipschitz state-feedback policy.
We also study output-error delay identification from a fixed
input--output record, deriving the loss gradient and
establishing projected-flow descent and local exponential
convergence under exact model matching and an excitation
condition. These identification results do not imply
stability of an online adaptive predictor.
Numerical experiments retained from the conference study
compare PID, LQR, and DDPG controllers on stable and open-loop
unstable first-order plants with valve stiction.
The results illustrate the benefit of predictor compensation
in the tested configurations and the sensitivity of standalone
DDPG to delay and plant changes.
\end{abstract}

\allowdisplaybreaks

\section{Introduction}
\label{sec:introduction}

Reinforcement learning provides a framework for improving feedback policies from observed transitions and rewards. Its connections with optimal and adaptive control have motivated both theoretical developments and practical applications \cite{Lewis2012,Kiumarsi2018,Jiang2015}. Deterministic policy gradients and deep deterministic policy gradient (DDPG) extend this framework to continuous actions, making actor--critic methods attractive for control problems in which discrete action sets are restrictive \cite{Silver2014,Lillicrap2016}. However, a policy trained on a nominal environment need not retain its performance or stability when the deployed system differs from the training model.

This distinction between learning performance and closed-loop guarantees is important under uncertainty and disturbances. Robust adaptive dynamic programming and disturbance-aware actor--critic methods incorporate additional structure into learning-based control \cite{Jiang2014,Song2016}, while quadratic-constraint methods provide tools for analyzing systems with fixed neural network controllers \cite{Yin2022}. Comparisons of reinforcement learning and adaptive control for uncertain linear systems also illustrate the different information requirements and robustness properties of the two approaches \cite{Wafi-SIAM,Wafi-SIAM-arXiv}. These results motivate examining the deployed feedback interconnection explicitly, rather than treating successful training as a stability certificate.

Time delays introduce a further deployment difficulty. Communication, computation, sensing, and actuation can make a control action depend on information that no longer represents the current plant condition. Networked-control studies document the resulting stability and performance limitations and develop conditions that account for communication constraints \cite{Hespanha2007,Zhang2001,Walsh2002}. Learning methods can incorporate delay information into an augmented state or history; for example, adaptive dynamic programming has been developed for discrete-time systems with multiple delays \cite{Zhang2020Delay}. Such approaches require a suitable representation of the delayed dynamics and may increase the dimension of the learning problem.

The Smith predictor offers a complementary approach: the controller is designed around nominal delay-free dynamics, while a model-based feedback construction compensates for a specified delay \cite{Smith1957,Normey2008}. Modifications address important limitations of the original construction, including integrating processes and robustness to model errors \cite{Astrom1994,Normey1999}. Predictor-feedback theory extends delay compensation to nonlinear systems \cite{Krstic2008,Krstic2010Nonlinear}, time-varying input delays \cite{Krstic2010Varying}, and distributed input delays \cite{Bekiaris2011}. The present study considers a constant input delay. These broader results provide context, but their guarantees do not automatically apply to a Smith predictor containing an arbitrary learned policy.

Model accuracy remains central to predictor-based control. Under perfect matching, the Smith-predictor algebra removes the delay from the nominal feedback denominator, while the physical output still exhibits the transport delay. Under mismatch, the same construction creates residual feedback dynamics that must be assessed. Small-gain and integral-quadratic-constraint methods provide general robustness tools \cite{Megretski1997}; the $\nu$-gap metric characterizes plant separation relevant to robustness transfer for a fixed linear controller \cite{Vinnicombe1993}. State-space methods offer a different route to time-delay stability analysis \cite{Fridman2002}. Applying these tools requires specifying the controller class, internal-stability assumptions, and the interconnection to which each condition refers.

An uncertain delay also raises an identification problem. Parametric identification uses a model and a data record to estimate unknown quantities under suitable excitation and regularity assumptions \cite{Ljung1978,Wafi-SysID-arXiv}. Delay-adaptive predictor feedback has been studied for linear and nonlinear systems with unknown actuator delays \cite{Bresch2010,Bresch2014}. Here, output-error delay identification is treated on a fixed input--output record. Descent of this identification loss and local convergence of its minimizer are distinguished from stability of a controller coupled to an online estimator.

Reliable model fitting and feedback also depend on the available measurements. Nonlinear filtering provides a framework for estimating quantities that are not measured directly \cite{Julier2004,Arasaratnam2009}. Related studies address satellite-tracking filters \cite{Wafi-AIP,Nardin2025,Wafi-AIP-arXiv} and consensus-based unscented estimation for vehicle dynamics \cite{Wafi-JRC}. Although these applications differ from delay compensation, they illustrate the role of estimation and measurement processing in model-based feedback. The present analysis uses the specified measured signals and nominal predictor model; it does not introduce a separate filtering subsystem.

This paper expands the Smith-predictor-assisted reinforcement learning study presented in \cite{Wafi-ACC24}. The policy is trained on a nominal delay-free environment and then deployed with the predictor, separating policy training from delay compensation. The expanded treatment makes the assumptions and limits of the analysis explicit through three main elements:
\begin{enumerate}
    \item Exact reference and disturbance maps are derived for the linear Smith-predictor interconnection under simultaneous plant and delay mismatch. Classical small-gain and $\nu$-gap results are then used to state sufficient robustness conditions for the relevant fixed linear controller.
    \item A delay-independent Razumikhin condition is given for a separate delayed state-feedback system with a fixed Lipschitz policy. An output-error delay-identification problem is also formulated, with frozen-data descent and local convergence under model-matching and finite-window excitation assumptions.
    \item The conference simulations are retained to compare predictor-assisted DDPG, standalone DDPG, PID, and LQR on stable and unstable first-order plants with valve stiction. Their observed behavior is interpreted separately from the sufficient theoretical conditions.
\end{enumerate}

Section~\ref{sec:problem} states the plant and control problem. Section~\ref{sec:smith} derives the Smith-predictor representation, and Section~\ref{sec:robustness} presents robustness and stability conditions. Section~\ref{sec:adaptation} develops output-error delay identification. Section~\ref{sec:drl} describes the DDPG policy and its deployment, Section~\ref{sec:numerics} presents the retained numerical examples, and Section~\ref{sec:conclusion} concludes the paper.

\section{Problem Formulation}
\label{sec:problem}

Consider the linear plant with a constant input delay
\begin{align}
    \begin{aligned}
        \dot x(t)&=Ax(t)+Bu(t-\ell_p),\\
        y(t)&=Cx(t)+Du(t-\ell_p)+d(t),
    \end{aligned}
    \label{eq:plant}
\end{align}
where $x(t)\in\R^n$, $u(t)\in\R^m$, and $y(t)\in\R^p$.
The matrices $A$, $B$, $C$, and $D$ have compatible dimensions,
$\ell_p\geq0$ is the plant delay, and $d(t)$ is an additive
output disturbance. The initial state $x(0)=x_0$ and an input
history $u(t)=u_h(t)$ for $t\in[-\ell_p,0)$ are prescribed.

Under zero initial conditions, the transfer function from
$u$ to $y-d$ is
\begin{equation}
    P(s)=G_p(s)T_p(s),\qquad
    G_p(s)=C(sI-A)^{-1}B+D,\qquad
    T_p(s)=e^{-\ell_p s}.
    \label{eq:plant-tf}
\end{equation}
The transfer-function analysis below focuses on SISO plants.

The Smith predictor uses the nominal model
\begin{equation}
    P_n(s)=G_n(s)T_n(s),\qquad
    T_n(s)=e^{-\ell_n s},
    \label{eq:nominal-model}
\end{equation}
where $G_n$ and the nominal delay $\ell_n\geq0$ are specified
and held fixed during deployment. We consider both perfect
matching, $G_n=G_p$ and $\ell_n=\ell_p$, and mismatch in the
plant dynamics or delay. No online delay estimation is assumed.

The controller is a deterministic policy
\begin{equation}
    u(t)=\pi_\theta(\mathbf{x}(t)),\qquad
    \mathbf{x}(t)=\operatorname{col}
    \bigl(y_f(t),e_f(t),e_I(t)\bigr),
    \label{eq:policy}
\end{equation}
where $y_f$ is the feedback signal supplied to the policy,
$e_f=r-y_f$, and $\dot e_I=e_f$. For standalone reinforcement
learning, $y_f=y$; for Smith-predictor-assisted control, $y_f$
is constructed from the measured output and nominal-model
outputs, as described in Section~\ref{sec:smith}.
The observation vector $\mathbf{x}$ is distinct from the
physical plant state $x$. The actor parameters $\theta$ are
held fixed after training.

Training seeks a policy that approximately minimizes
\begin{equation}
    J_T(\theta)=
    \mathbb{E}\!\left[
        \int_0^T
        c\bigl(\mathbf{x}(t),
        \pi_\theta(\mathbf{x}(t)),t\bigr)\,dt
    \right]
    \label{eq:training-objective}
\end{equation}
on a nominal delay-free environment, where $T>0$ is the
training horizon and the expectation is taken over the
training randomness. This objective alone does not guarantee
closed-loop stability during deployment.

Our objective is to characterize the effect of plant and
delay mismatch on the Smith-predictor architecture and to
identify sufficient conditions for closed-loop stability.
The theoretical analysis distinguishes results for linear
controllers from those applicable to a fixed nonlinear policy.

\section{Smith-Predictor Closed-Loop Representation}
\label{sec:smith}

The Smith-predictor feedback signal is
\begin{equation}
    y_f=y-y_n+y_0,
    \label{eq:smith-signal}
\end{equation}
where $y_n=G_nT_nu$ and $y_0=G_nu$ are the outputs of the nominal delayed
and delay-free models, respectively. Under zero initial
conditions, $Y_n=G_nT_nU$ and $Y_0=G_nU$.
Consequently,
\begin{equation}
    Y_f=Y+G_n(1-T_n)U.
    \label{eq:predictor-feedback}
\end{equation}

For the transfer-function analysis, consider a SISO linear
time-invariant controller $K(s)$ with
\begin{equation}
    U=K(R-Y_f).
    \label{eq:linear-controller}
\end{equation}
These expressions also describe the linearized interconnection
around an equilibrium when the deployed policy and its
observation dynamics admit a realization of this form.
In that case, the conclusions concern the linearized system.

\begin{figure}[t!]
    \centering
    \resizebox{.5\linewidth}{!}{%
    \begin{tikzpicture}[
        >=Latex,
        block/.style={
            draw, rectangle,
            minimum width=1.05cm,
            minimum height=0.85cm,
            fill=white
        },
        sum/.style={
            draw, circle,
            minimum size=0.42cm,
            inner sep=0pt,
            fill=white
        },
        signal/.style={->, thick},
        wire/.style={thick},
        every node/.style={font=\small}
    ]
    
    \node (ref) at (-1,0) {$r$};
    \node[sum] (error) at (0,0) {};
    \node[block] (controller) at (1.3,0) {$K$};
    \coordinate (branch) at (2.2,0);
    \node[block] (nominal) at (3.3,0) {$G_n$};
    \coordinate (modelbranch) at (4.25,0);
    \node[block] (delay) at (5.25,0) {$T_n$};
    
    \node[sum] (inputsum) at (3.05,2) {};
    \node[block] (plant) at (4.35,2) {$P$};
    \node[sum] (outputsum) at (5.65,2) {};
    \coordinate (outputbranch) at (6.65,2);
    \node (output) at (7.4,2) {$y$};
    
    \node[sum] (residual) at (6.65,0) {};
    \node[sum] (feedbacksum) at (4.25,-1.5) {};
    
    \node (inputdist) at (3.05,3) {$d_1$};
    \node (outputdist) at (5.65,3) {$d$};
    
    \draw[signal] (ref) -- (error);
    \draw[signal] (error) -- node[above] {$e_f$} (controller);
    \draw[wire] (controller) -- node[above] {$u$} (branch);
    \draw[signal] (branch) -- (nominal);
    
    \draw[signal] (branch) |- (inputsum);
    \draw[signal] (inputdist) -- (inputsum);
    \draw[signal] (inputsum) -- (plant);
    \draw[signal] (plant) -- (outputsum);
    \draw[signal] (outputdist) -- (outputsum);
    \draw[wire] (outputsum) -- (outputbranch);
    \draw[signal] (outputbranch) -- (output);
    
    \draw[wire] (nominal) -- (modelbranch);
    \draw[signal] (modelbranch) --
        node[above] {$y_0$} (delay);
    \draw[signal] (delay) --
        node[above] {$y_n$} (residual);
    
    \draw[signal] (outputbranch) -- (residual);
    \draw[signal] (modelbranch) -- (feedbacksum);
    \draw[signal] (residual) |- 
        node[pos=0.75,below] {$y-y_n$} (feedbacksum);
    \draw[signal] (feedbacksum) -|
        node[pos=0.35,below] {$y_f$} (error);
    
    \node at ($(error)+(-0.33,0.18)$) {$+$};
    \node at ($(error)+(0.18,-0.33)$) {$-$};
    
    \node at ($(inputsum)+(-0.33,0.18)$) {$+$};
    \node at ($(inputsum)+(0.18,0.33)$) {$+$};
    
    \node at ($(outputsum)+(-0.33,0.18)$) {$+$};
    \node at ($(outputsum)+(0.18,0.33)$) {$+$};
    
    \node at ($(residual)+(-0.33,0.18)$) {$-$};
    \node at ($(residual)+(0.18,0.33)$) {$+$};
    
    \node at ($(feedbacksum)+(0.18,0.33)$) {$+$};
    \node at ($(feedbacksum)+(0.33,-0.18)$) {$+$};
    
    \fill (branch) circle (1.5pt);
    \fill (modelbranch) circle (1.5pt);
    \fill (outputbranch) circle (1.5pt);
    
    \begin{scope}[on background layer]
        \node[
            draw, dashed, rounded corners,
            fill=gray!6,
            fit=(inputsum)(plant)(outputsum)(outputbranch),
            inner sep=0.3cm
        ] (processbox) {};
    
        \node[
            draw, dashed, rounded corners,
            fill=gray!6,
            fit=(nominal)(delay)(residual)(feedbacksum),
            inner sep=0.3cm
        ] (predictorbox) {};
    \end{scope}
    
    \node[anchor=south east] at (processbox.north east)
        {Process};
    \node[anchor=north west] at (predictorbox.south west)
        {Smith predictor};
    
    \end{tikzpicture}%
    }
    \caption{Smith-predictor architecture with nominal models
    $G_n$ and $T_n$. The controller receives the corrected
    feedback $y_f=y-y_n+y_0$. The input disturbance $d_1$
    is set to zero in the present analysis.}
    \label{F1}
\end{figure}
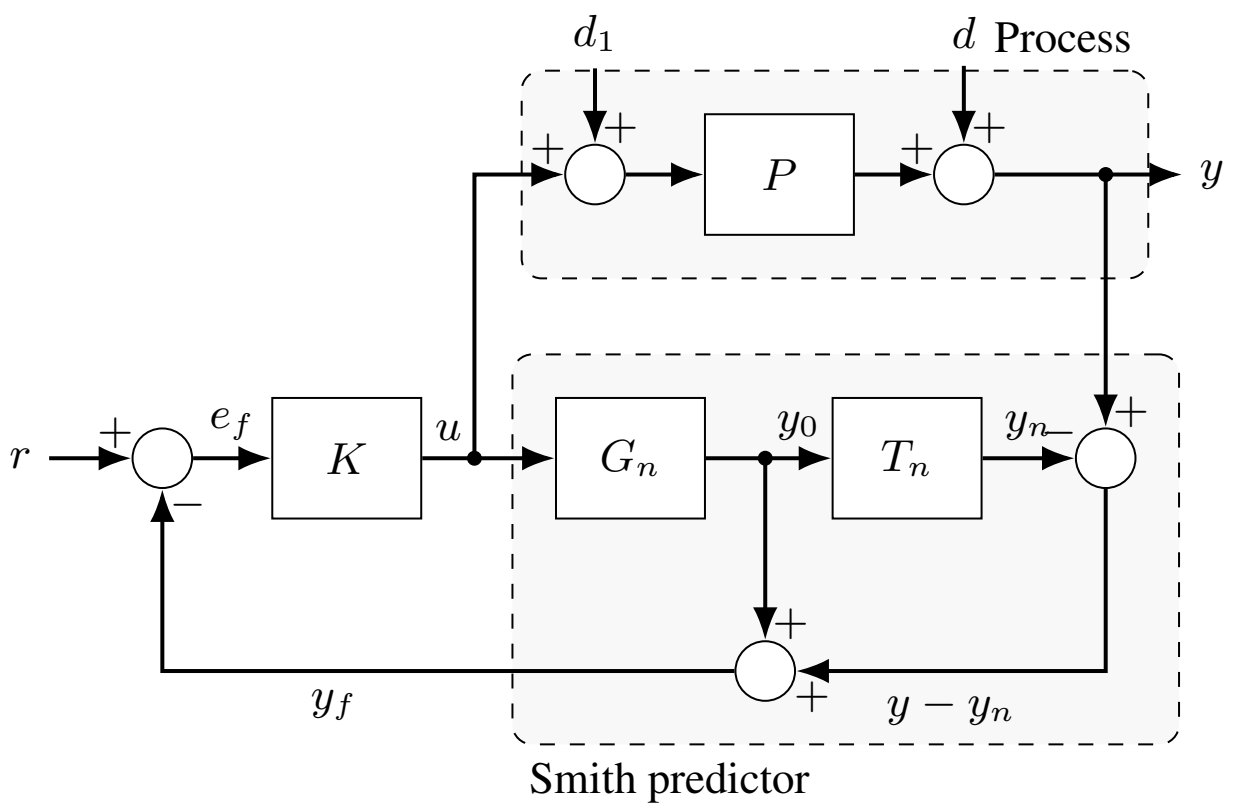

Define
\begin{equation}
    \Phi_n(s)=1+K(s)G_n(s)\bigl(1-T_n(s)\bigr).
    \label{eq:phi}
\end{equation}

\begin{proposition}[Smith-predictor maps]
Suppose the linear interconnection is well posed and all
initial conditions are zero. Let $D_o(s)$ denote the Laplace
transform of the additive output disturbance $d(t)$.
The reference-to-output map, with $D_o=0$, and the
disturbance-to-output map, with $R=0$, are
\begin{equation}
    \begin{aligned}
        \frac{Y}{R}=\frac{KP}{\Phi_n+KP},
        \qquad
        \frac{Y}{D_o}=\frac{\Phi_n}{\Phi_n+KP},
    \end{aligned}
    \label{eq:smith-maps}
\end{equation}
where $P=G_pT_p$.

Under perfect matching, $G_n=G_p=:G$ and $T_n=T_p=:T$,
these become
\begin{equation}
    \begin{aligned}
        \frac{Y}{R}=\frac{KG}{1+KG}\,T,
        \qquad
        \frac{Y}{D_o}=\frac{1+KG(1-T)}{1+KG}.
    \end{aligned}
    \label{eq:perfect-maps}
\end{equation}
Thus, the pure delay disappears from the nominal feedback
denominator but remains in the physical reference response.
\end{proposition}

\begin{proof}
Substituting \eqref{eq:predictor-feedback} into
\eqref{eq:linear-controller} yields $\Phi_nU=K(R-Y)$.
Using $Y=PU+D_o$, we obtain
\[
    (\Phi_n+KP)U=K(R-D_o),
\]
and hence
\[
    Y=\frac{KP}{\Phi_n+KP}R
      +\frac{\Phi_n}{\Phi_n+KP}D_o,
\]
which proves \eqref{eq:smith-maps}. Under perfect matching,
$P=GT$ and
\[
    \Phi_n+KP=1+KG(1-T)+KGT=1+KG.
\]
Substitution gives \eqref{eq:perfect-maps}.
\end{proof}

To describe mismatch, define
\begin{equation}
    E(s):=G_p(s)T_p(s)-G_n(s)T_n(s).
    \label{eq:model-error}
\end{equation}
The predictor feedback then satisfies
\begin{equation}
    Y_f=(G_n+E)U+D_o.
    \label{eq:effective-plant}
\end{equation}
Thus, from the controller's perspective, mismatch introduces
the additive model error $E$ into the nominal delay-free plant.
The feedback denominator is
\begin{equation}
    \chi(s):=\Phi_n(s)+K(s)P(s)
    =1+KG_n+KE.
    \label{eq:characteristic}
\end{equation}
Define the nominal sensitivity and complementary sensitivity
functions
\begin{equation}
    S_0:=\frac{1}{1+KG_n},\qquad
    \Phi_0:=\frac{KG_n}{1+KG_n}.
    \label{eq:nominal-sensitivities}
\end{equation}
Then the following identity holds wherever the expressions
are defined:
\begin{equation}
    \chi=(1+KG_n)(1+KS_0E).
    \label{eq:characteristic-factor}
\end{equation}
This representation separates the nominal feedback
denominator from the residual mismatch factor.

If the plant mismatch admits the multiplicative description
$G_p=G_n(1+\Delta_G)$, then
\begin{equation}
    KS_0E
    =\Phi_0\bigl((1+\Delta_G)T_p-T_n\bigr).
    \label{eq:multiplicative-residual}
\end{equation}
In particular, for plant mismatch alone, $T_p=T_n=:T$,
\begin{equation}
    KS_0E=\Phi_0T\Delta_G,
    \label{eq:plant-mismatch-residual}
\end{equation}
whereas for delay mismatch alone, $G_p=G_n$,
\begin{equation}
    KS_0E=\Phi_0(T_p-T_n).
    \label{eq:delay-mismatch-residual}
\end{equation}

\begin{remark}[Internal stability]
The transfer maps above describe zero-initial-condition
input--output behavior. Stability of these maps alone does
not establish internal stability of the complete realization,
because unstable internal modes may be hidden by cancellation.
Accordingly, internal-stability conclusions require additional
assumptions on the plant, predictor, and controller realizations.
This distinction is especially important for open-loop
unstable plants.
\end{remark}

\begin{remark}[Delay approximations]
All identities above use the exact delay operator
$e^{-\ell s}$. Finite-order Pad\'e or truncated Taylor
approximations change the model and may yield different
stability boundaries. Routh--Hurwitz tests applied to these
approximations establish stability of the approximate
finite-dimensional model, rather than the exact delay system.
\end{remark}

\section{Robustness and Stability Conditions}
\label{sec:robustness}

\subsection{Small-gain condition}

Recall the model error $E=G_pT_p-G_nT_n$ and the nominal
sensitivity $S_0=(1+KG_n)^{-1}$. The residual mismatch loop is
\begin{equation}
    M:=KS_0E.
    \label{eq:mismatch-loop}
\end{equation}
The following result gives a sufficient stability condition
for stable plant and predictor realizations.

\begin{theorem}[Robust stability under joint mismatch]
Consider a well-posed SISO linear Smith-predictor
interconnection. Assume that the delay-free plant and
nominal-model realizations are internally stable, the delays
are nonnegative constants, and $K$ internally stabilizes $G_n$.
If
\begin{equation}
    \norm{M}_\infty<1,
    \label{eq:small-gain}
\end{equation}
then the complete Smith-predictor interconnection is
internally stable.
\end{theorem}

\begin{proof}
Nominal internal stability implies that $S_0$, $KS_0$,
and $G_nS_0$ are stable. Since the plant and nominal model
have stable realizations, $E$ is stable, and hence so is $M$.
The small-gain condition guarantees that
\[
    H:=(1+M)^{-1}
\]
is stable. The reference and output-disturbance responses
satisfy
\[
    U=HKS_0(R-D_o),\qquad Y=PU+D_o.
\]
More generally, for additive signals $v$ at the controller
output and $w$ at its feedback input,
\[
    U=H\bigl[KS_0(R-D_o-w)+S_0v\bigr].
\]
Thus, the controller input and output remain stable under
independent disturbances. The plant and predictor signals
are generated by stable model blocks driven by these signals.
Together with nominal internal stability, this establishes
internal stability of the full interconnection.
\end{proof}

\begin{remark}
The assumption of stable plant and predictor realizations
excludes hidden unstable modes. The theorem therefore does
not certify the open-loop unstable examples. Such cases
require an analysis of the complete realization rather than
the feedback denominator alone.
\end{remark}

If $G_p=G_n(1+\Delta_G)$, then
\begin{equation}
    M=\Phi_0\bigl((1+\Delta_G)T_p-T_n\bigr),
    \qquad
    \Phi_0=\frac{KG_n}{1+KG_n}.
    \label{eq:multiplicative-mismatch-loop}
\end{equation}
For plant mismatch alone, $T_p=T_n=:T$, the condition becomes
\begin{equation}
    \norm{\Phi_0T\Delta_G}_\infty
    =\norm{\Phi_0\Delta_G}_\infty<1,
    \label{eq:plant-smallgain}
\end{equation}
because $|T(j\omega)|=1$. For delay mismatch alone,
$G_p=G_n$, it becomes
\begin{equation}
    \norm{\Phi_0(T_p-T_n)}_\infty<1.
    \label{eq:pure-delay-smallgain}
\end{equation}
In particular,
\begin{equation}
    |T_p(j\omega)-T_n(j\omega)|
    =2\left|\sin\!\left(
        \frac{\omega(\ell_p-\ell_n)}{2}
    \right)\right|.
    \label{eq:delay-difference}
\end{equation}
Thus, stability depends on the delay mismatch weighted by
the nominal complementary sensitivity over all frequencies.

\begin{corollary}[Sufficient delay-mismatch bound]
Under the assumptions of the theorem, suppose $G_p=G_n$
and $s\Phi_0(s)$ is stable and proper. If
\begin{equation}
    |\ell_p-\ell_n|\,\norm{s\Phi_0}_\infty<1,
    \label{eq:delay-mismatch-bound}
\end{equation}
then the Smith-predictor interconnection is internally stable.
\end{corollary}

\begin{proof}
Using $2|\sin(a/2)|\leq |a|$ gives
\[
    \norm{\Phi_0(T_p-T_n)}_\infty
    \leq |\ell_p-\ell_n|\,\norm{s\Phi_0}_\infty.
\]
The conclusion follows from \eqref{eq:small-gain}.
\end{proof}

\subsection{$\nu$-gap robustness transfer}

The $\nu$-gap provides a complementary robustness test
for a fixed linear controller. Consider SISO plants
$P_i=N_iD_i^{-1}$ with normalized coprime factorizations,
and define
\begin{equation}
    \Phi_{21}=N_2^\sim N_1+D_2^\sim D_1,
    \qquad
    \Psi_{21}=-D_2N_1+N_2D_1,
    \label{eq:gap-factors}
\end{equation}
where $H^\sim(s)=H^\top(-s)$ for real transfer functions.
Within the standard coprime-factor framework, the $\nu$-gap is
\begin{equation}
    \delta_\nu(P_1,P_2)=
    \begin{cases}
        \norm{\Psi_{21}}_\infty,
        & \Phi_{21}\neq0
          \text{ on the Nyquist contour},\\[-0.2em]
        & \operatorname{wno}(\Phi_{21})=0,\\
        1,&\text{otherwise}.
    \end{cases}
    \label{eq:nugap}
\end{equation}
The contour condition includes the point at infinity.

For a linear controller $K$ internally stabilizing $P_1$,
define its generalized stability margin
\begin{equation}
    b_{P_1,K}:=
    \left\|
        \begin{bmatrix}I\\K\end{bmatrix}
        (I+P_1K)^{-1}
        \begin{bmatrix}I&P_1\end{bmatrix}
    \right\|_\infty^{-1}.
    \label{eq:margin}
\end{equation}

\begin{theorem}[$\nu$-gap robust-stability test]
If $K$ internally stabilizes $P_1$ and
\begin{equation}
    \delta_\nu(P_1,P_2)<b_{P_1,K},
    \label{eq:nugap-test}
\end{equation}
then $K$ internally stabilizes $P_2$. Moreover,
\begin{equation}
    \arcsin b_{P_2,K}
    \geq
    \arcsin b_{P_1,K}
    -\arcsin\delta_\nu(P_1,P_2).
    \label{eq:gap-margin-bound}
\end{equation}
\end{theorem}

This is the standard $\nu$-gap robustness theorem
\cite{B2}. Its application requires the same controller
in both feedback interconnections.

For the Smith-predictor architecture, one may compare the
effective plants
\begin{equation}
    P_{\mathrm{eff},1}=G_n,\qquad
    P_{\mathrm{eff},2}=G_n+E,
    \label{eq:effective-gap-plants}
\end{equation}
with the controller $K$. Under the stable-realization
assumptions of the preceding theorem, stability of the
effective feedback interconnection also establishes stability
of the complete Smith-predictor architecture.

Alternatively, when comparing physical plants $P_1$ and
$P_2$, the controller in the gap test must be the fixed
composite compensator
\begin{equation}
    K_{\mathrm{SP}}=\frac{K}{\Phi_n},
    \label{eq:composite-controller}
\end{equation}
and nominal internal stability of its full realization
must be verified.

For a nonlinear neural policy, the linear margin
\eqref{eq:margin} cannot be evaluated by directly substituting
the network for $K$. It may instead be used for a justified
linearization around an equilibrium. Accordingly, the
$\nu$-gap values reported in Section~\ref{sec:numerics}
quantify model separation; a stability certificate additionally
requires the corresponding controller margin and the strict
inequality \eqref{eq:nugap-test}. If finite-order delay
approximations are used, the certificate concerns those
approximate models.

\subsection{Delay-independent condition for state feedback}

We next consider a separate state-feedback configuration,
without the Smith predictor or integral observation dynamics:
\begin{equation}
    \dot x(t)=Ax(t)+B\pi_\theta(x(t-\ell_p)).
    \label{eq:delayed-state-feedback}
\end{equation}
This condition applies directly to a fixed nonlinear policy,
but does not certify the full architecture of
Section~\ref{sec:smith}.

\begin{assumption}
The reference and disturbances are zero,
$\pi_\theta(0)=0$, and the policy
$\pi_\theta:\R^n\to\R^m$ is globally Lipschitz:
\begin{equation}
    \norm{\pi_\theta(z_1)-\pi_\theta(z_2)}
    \leq L_\pi\norm{z_1-z_2}.
    \label{eq:lipschitz}
\end{equation}
\end{assumption}

\begin{proposition}[Razumikhin sufficient condition]
Suppose there exist $Q=Q^\top\succ0$, $q>1$, and $\alpha>0$.
Define $c_Q=qL_\pi^2/\lambda_{\min}(Q)$. If
\begin{equation}
    \begin{bmatrix}
        A^\top Q+QA+\alpha c_QQ & QB\\
        B^\top Q & -\alpha I
    \end{bmatrix}\prec0,
    \label{eq:razumikhin-lmi}
\end{equation}
then the origin of \eqref{eq:delayed-state-feedback}
is globally asymptotically stable for every finite
constant delay $\ell_p\geq0$.
\end{proposition}

\begin{proof}
Let $V(x)=x^\top Qx$. Under the Razumikhin condition
\begin{equation}
    V(x(t+\sigma))\leq qV(x(t)),
    \qquad \sigma\in[-\ell_p,0],
    \label{eq:raz-condition}
\end{equation}
the delayed control $v=\pi_\theta(x(t-\ell_p))$ satisfies
\[
    \norm{v}^2
    \leq L_\pi^2\norm{x(t-\ell_p)}^2
    \leq c_QV(x(t)).
\]
Consequently,
\[
    \dot V
    \leq \dot V+\alpha(c_Qx^\top Qx-v^\top v)
    =
    \begin{bmatrix}x\\v\end{bmatrix}^{\!\top}
    \begin{bmatrix}
        A^\top Q+QA+\alpha c_QQ & QB\\
        B^\top Q & -\alpha I
    \end{bmatrix}
    \begin{bmatrix}x\\v\end{bmatrix}.
\]
Strict negativity gives $\dot V\leq-\varepsilon\norm{x}^2$
for some $\varepsilon>0$ whenever
\eqref{eq:raz-condition} holds.
The Lyapunov--Razumikhin theorem establishes the result.
\end{proof}

\begin{remark}
Condition \eqref{eq:razumikhin-lmi} is delay independent
and conservative. In particular, it requires $A$ to be
Hurwitz and therefore does not address stabilization of
open-loop unstable plants. It is an a posteriori test:
DDPG training alone does not enforce this condition.
Because $c_Q$ depends on $Q$, the displayed condition is
a matrix inequality, rather than an LMI jointly in all
the stated variables.
\end{remark}

\section{Output-Error Delay Identification}
\label{sec:adaptation}

We consider identification of an unknown constant plant delay
by minimizing the nominal model's output prediction error
over a fixed input--output record. We derive the loss gradient
and establish descent and local convergence properties under
suitable regularity and excitation conditions.

Let $\widehat y(t;\ell_n)$ denote the nominal-model output
for a candidate delay $\ell_n$, driven by the recorded input
$u$ with a specified initialization and input history.
Over a fixed window $[t_k,t_k+T_w]$, define
\begin{equation}
    \mathcal{J}_k(\ell_n)=\frac12
    \int_{t_k}^{t_k+T_w}
    \norm{y(t)-\widehat y(t;\ell_n)}^2\,dt.
    \label{eq:delay-loss}
\end{equation}
Differentiation is with respect to the scalar parameter
$\ell_n$, while the recorded signals $u$ and $y$ are held fixed.

For the nominal transfer function $P_n(s,\ell_n)=G_n(s)e^{-\ell_ns}$,
\begin{equation}
    \frac{\partial P_n}{\partial\ell_n}(s,\ell_n)
    =-sG_n(s)e^{-\ell_ns}=-sP_n(s,\ell_n).
    \label{eq:delay-transfer-sensitivity}
\end{equation}
Hence the open-loop output sensitivity $z(t;\ell_n):=\partial\widehat y/\partial\ell_n$ is generated by
\begin{equation}
    Z(s;\ell_n)=-sP_n(s,\ell_n)U(s).
    \label{eq:sensitivity-filter}
\end{equation}
When the model is simulated inside the closed loop, $z$ must instead be obtained by differentiating the complete predictor equations, because $u$ may depend on $\ell_n$ through feedback.

\begin{lemma}[Exact loss gradient]
If $\widehat y(t;\ell_n)$ is differentiable in $\ell_n$ and differentiation can be interchanged with integration, then
\begin{equation}
    \mathcal{J}'_k(\ell_n)
    =-\int_{t_k}^{t_k+T_w}
    \bigl(y(t)-\widehat y(t;\ell_n)\bigr)^\top
    z(t;\ell_n)\,dt.
    \label{eq:loss-gradient}
\end{equation}
\end{lemma}

Let $\mathcal{L}=[\ell_{\min},\ell_{\max}]$ be a known physical interval. The batch update is
\begin{equation}
    \ell_{n,k+1}=\Pi_{\mathcal{L}}
    \left(\ell_{n,k}-\gamma_k\mathcal{J}'_k(\ell_{n,k})\right),
    \qquad \gamma_k>0,
    \label{eq:projected-update}
\end{equation}
where $\Pi_{\mathcal L}$ denotes Euclidean projection. The continuous-time analogue is the projected flow
\begin{equation}
    \dot\ell_n=\Pi_{T_{\mathcal L}(\ell_n)}
    \left(-\gamma\mathcal J'(\ell_n)\right).
    \label{eq:projected-flow}
\end{equation}

\begin{proposition}[Descent of the frozen-data loss]
Assume a fixed data window, $\mathcal{J}\in C^1(\mathcal L)$, and use \eqref{eq:projected-flow}. Then $\mathcal L$ is forward invariant and
\begin{equation}
    \dot{\mathcal J}(\ell_n)
    =\mathcal J'(\ell_n)\dot\ell_n\leq0.
    \label{eq:loss-descent}
\end{equation}
Every accumulation point is a projected stationary point of $\mathcal J$.
\end{proposition}

\begin{proof}
Projection onto the tangent cone makes $\mathcal L$ invariant. The defining variational inequality of Euclidean projection gives
$\mathcal J'(\ell_n)\Pi_{T_{\mathcal L}}(-\gamma\mathcal J'(\ell_n))\leq0$.
LaSalle's invariance principle for projected systems gives the final claim.
\end{proof}

\begin{theorem}[Local delay convergence]
Suppose the plant belongs to the nominal family, so that $y(t)=\widehat y(t;\ell_p)$ on the identification window, and $\ell_p\in\operatorname{int}\mathcal L$. Assume $\widehat y$ is twice continuously differentiable with respect to $\ell$ near $\ell_p$ and
\begin{equation}
    \int_{t_k}^{t_k+T_w}\norm{z(t;\ell_p)}^2dt\geq\mu>0.
    \label{eq:identifiability}
\end{equation}
Then $\ell_p$ is a strict local minimizer of $\mathcal J_k$. Moreover, there is a neighborhood of $\ell_p$ in which the unprojected gradient flow $\dot\ell_n=-\gamma\mathcal J'_k(\ell_n)$ converges exponentially to $\ell_p$.
\end{theorem}

\begin{proof}
At the true delay the prediction error is zero, so $\mathcal J'_k(\ell_p)=0$. Differentiating \eqref{eq:loss-gradient} once more gives
\begin{equation}
    \mathcal J''_k(\ell_p)=
    \int_{t_k}^{t_k+T_w}\norm{z(t;\ell_p)}^2dt\geq\mu.
\end{equation}
Continuity implies local strong convexity. Standard gradient-flow estimates then give exponential convergence.
\end{proof}

The excitation condition \eqref{eq:identifiability} ensures
local distinguishability of nearby delays on the recorded
window. Under model mismatch, measurement noise, or
initialization error, the true delay need not minimize
the prediction loss. The results above concern optimization
of a fixed-data loss and do not establish stability of a
Smith-predictor loop with an online-varying delay estimate.
Such an extension requires a separate analysis of the
coupled controller and estimator dynamics.

\section{DDPG Policy in the Smith-Predictor Architecture}
\label{sec:drl}

The controller is trained using deep deterministic policy
gradient (DDPG), with an actor $\pi_\theta$, a critic
$Q_\eta$, and target networks $\pi_{\theta^-}$ and
$Q_{\eta^-}$. The actor maps the observation vector
$\mathbf{x}\in\mathcal S$ to a continuous action
$u\in\mathcal A$.

With $\phi_0=\mathbf{x}$, the actor is represented by
\begin{equation}
    \begin{aligned}
        v_j&=W_j\phi_{j-1}+b_j,
        &\phi_j&=\sigma_j(v_j),
        \quad j=1,\ldots,L_a,\\
        \pi_\theta(\mathbf{x})
        &=W_{L_a+1}\phi_{L_a}+b_{L_a+1},
    \end{aligned}
    \label{eq:actor}
\end{equation}
where $W_j$, $b_j$, and $\sigma_j$ denote the layer weights,
biases, and activation functions. The critic processes
observations and actions through separate feedforward
branches, which are merged before producing the scalar
estimate $Q_\eta(\mathbf{x},u)$.

Training uses transitions
$(\mathbf{x}_i,u_i,\varrho_i,\mathbf{x}_i')$ sampled from
a replay buffer, where $\varrho_i$ denotes the reward and
$\mathbf{x}_i'$ the next observation. For a batch of
$N_b$ samples, the target value is
\begin{equation}
    q_i^{\mathrm{tar}}
    =\varrho_i+\gamma_d(1-\delta_i)
    Q_{\eta^-}\bigl(
        \mathbf{x}_i',
        \pi_{\theta^-}(\mathbf{x}_i')
    \bigr),
    \label{eq:ddpg-target}
\end{equation}
where $\gamma_d\in[0,1)$ is the discount factor and
$\delta_i=1$ for a terminal transition, with $\delta_i=0$
otherwise. The critic minimizes
\begin{equation}
    L(\eta)=\frac{1}{N_b}\sum_{i=1}^{N_b}
    \bigl(
        Q_\eta(\mathbf{x}_i,u_i)-q_i^{\mathrm{tar}}
    \bigr)^2,
    \label{eq:critic-loss}
\end{equation}
with the target values held fixed during each update.

The actor maximizes the batch objective
\begin{equation}
    \mathcal R_b(\theta)
    =\frac{1}{N_b}\sum_{i=1}^{N_b}
    Q_\eta\bigl(\mathbf{x}_i,\pi_\theta(\mathbf{x}_i)\bigr).
    \label{eq:actor-objective}
\end{equation}
With the critic parameters held fixed, its gradient is
\begin{equation}
    \nabla_\theta\mathcal R_b
    =\frac{1}{N_b}\sum_{i=1}^{N_b}
    \left(
        \frac{\partial\pi_\theta(\mathbf{x}_i)}
             {\partial\theta}
    \right)^{\!\top}
    \left.
        \nabla_uQ_\eta(\mathbf{x}_i,u)
    \right|_{u=\pi_\theta(\mathbf{x}_i)}.
    \label{eq:actor-gradient}
\end{equation}
Thus, the critic uses gradient descent on $L$, while the
actor uses gradient ascent on $\mathcal R_b$.

The target networks are initialized by copying the actor
and critic parameters and are subsequently updated through
\begin{equation}
    \eta^-\leftarrow\rho\eta^-+(1-\rho)\eta,
    \qquad
    \theta^-\leftarrow\rho\theta^-+(1-\rho)\theta,
    \label{eq:polyak}
\end{equation}
where $0<\rho<1$. The target actor supplies the continuous
action in \eqref{eq:ddpg-target}; no explicit maximization
over actions is required.

The conference study uses the tracking reward
\begin{equation}
    \varrho(e)=
    \begin{cases}
        10,& |e|\leq0.1,\\
        1/|e|,& |e|>0.1,
    \end{cases}
    \label{eq:tracking-reward}
\end{equation}
where $e$ is the tracking error supplied to the training
reward calculation. DDPG maximizes the discounted reward;
it therefore provides a reward-based training formulation
of the control objective in Section~\ref{sec:problem},
rather than an exact solution of the finite-horizon
optimization problem.

The actor is trained on the nominal delay-free environment
and its parameters are held fixed during deployment.
For standalone DDPG, the observations use the measured
output. In the Smith-predictor architecture, they use
the corrected feedback $y_f$ and its associated error
and integral, as defined in \eqref{eq:policy}.
The nominal predictor delay $\ell_n$ is also held fixed.
The identification procedure in Section~\ref{sec:adaptation}
is studied separately and is not used to update the
predictor in the retained numerical experiments.

DDPG training alone does not guarantee closed-loop stability.
The linear robustness tests in Section~\ref{sec:robustness}
apply to linear controllers or justified linearizations,
whereas the nonlinear state-feedback condition applies only
to the configuration specified in
\eqref{eq:delayed-state-feedback}.

\section{Numerical Examples and Interpretation}
\label{sec:numerics}

We retain the numerical experiments of the conference study,
which compare PID, LQR, and DDPG control under different
plant delays, with and without Smith-predictor compensation.
The stable delay-free plant is
\begin{equation}
    G_p(s)=\frac{k}{\tau s+1},
    \qquad
    k=3.8163,\quad \tau=156.46,
    \label{eq:first-order}
\end{equation}
and the delayed plant is $P(s)=G_p(s)e^{-\ell_p s}$.
The benchmark parameters in \cite{RB1} include a delay
of $2.5\,\mathrm{s}$. The deployment experiments vary
the delay and also consider an open-loop unstable variant,
obtained by changing the sign of the real plant pole.
The DDPG actor is trained only on the stable nominal
environment and is held fixed during deployment.

The actuator includes valve stiction. With commanded input
$u_j$, valve position $\varphi_j$, and friction parameters
$0\leq f_d\leq f_s$, the discrete-time model is
\begin{equation}
    \varphi_j=
    \begin{cases}
        u_j-f_d,
        &u_j-\varphi_{j-1}>f_s,\\
        u_j+f_d,
        &u_j-\varphi_{j-1}<-f_s,\\
        \varphi_{j-1},
        &|u_j-\varphi_{j-1}|\leq f_s.
    \end{cases}
    \label{eq:stiction}
\end{equation}
Here, $f_s$ and $f_d$ represent static and dynamic friction,
respectively. The valve remains stationary until the
commanded displacement exceeds the static-friction threshold.
This actuator nonlinearity is included in the simulations
but is not covered by the linear transfer-function analysis.

\begin{figure*}[t!]
    \centering
    \subfloat[
        \label{Ga}\scriptsize
        Unstable with SP, $\ell_p=10\,\mathrm{s}$
    ]{\includegraphics[width=.23\linewidth]{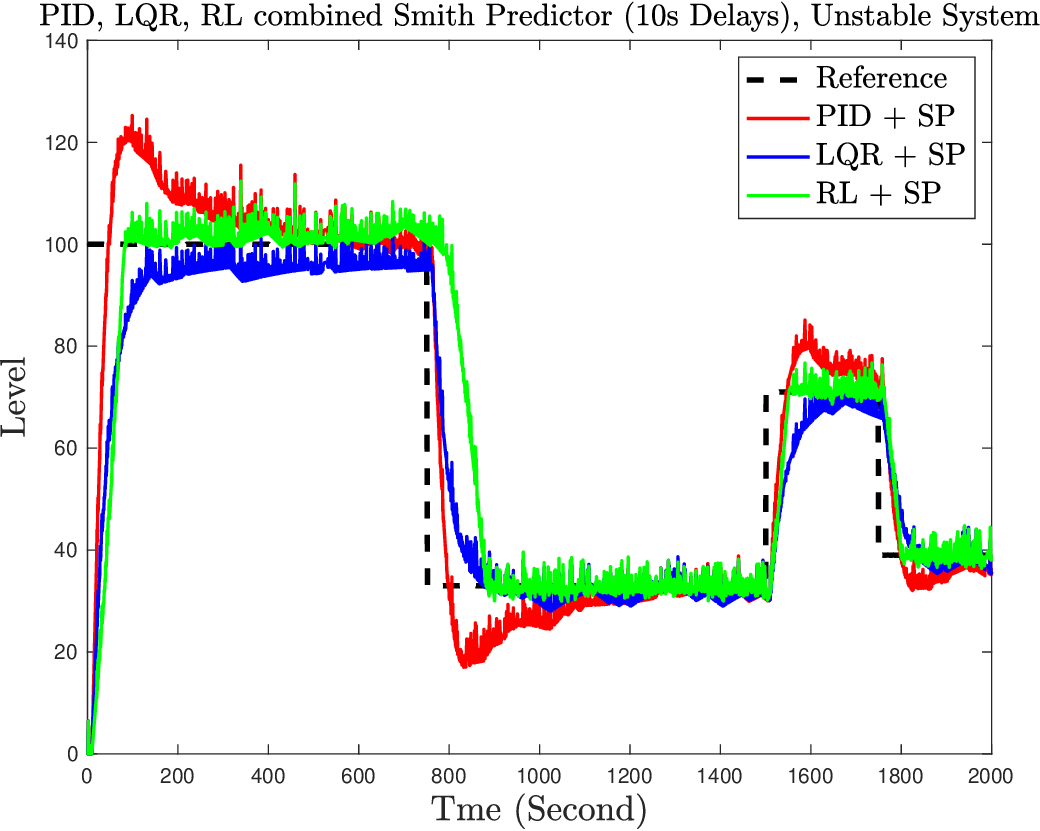}}\;
    \subfloat[
        \label{Gb}\scriptsize
        Unstable with SP, $\ell_p=25\,\mathrm{s}$
    ]{\includegraphics[width=.23\linewidth]{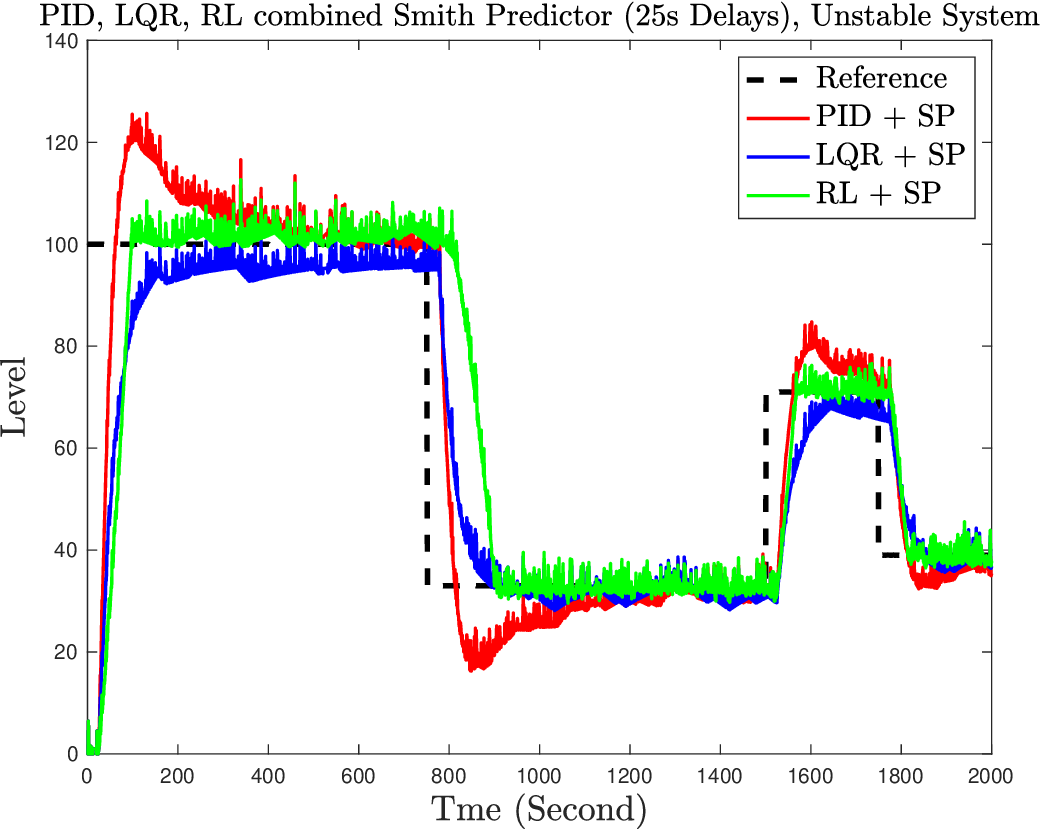}}\;
    \subfloat[
        \label{Gc}\scriptsize
        Unstable with SP, $\ell_p=100\,\mathrm{s}$
    ]{\includegraphics[width=.23\linewidth]{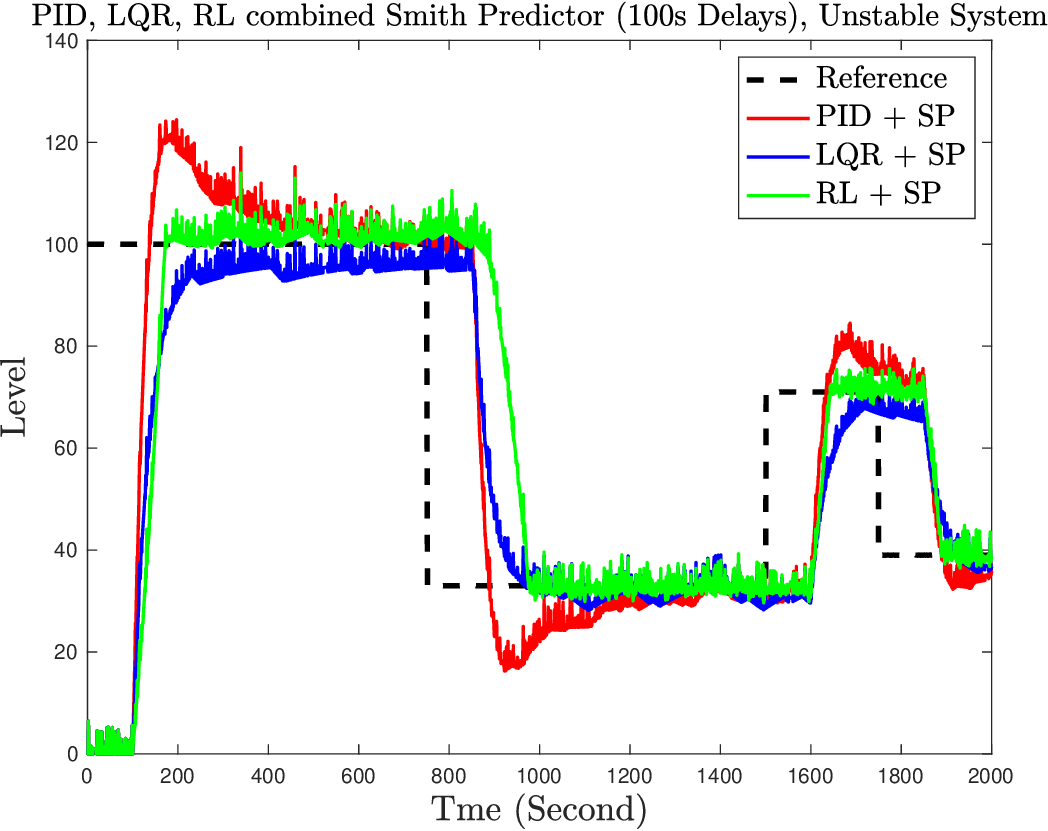}}\;
    \subfloat[
        \label{Gd}\scriptsize
        Unstable with SP, $\ell_p=150\,\mathrm{s}$
    ]{\includegraphics[width=.23\linewidth]{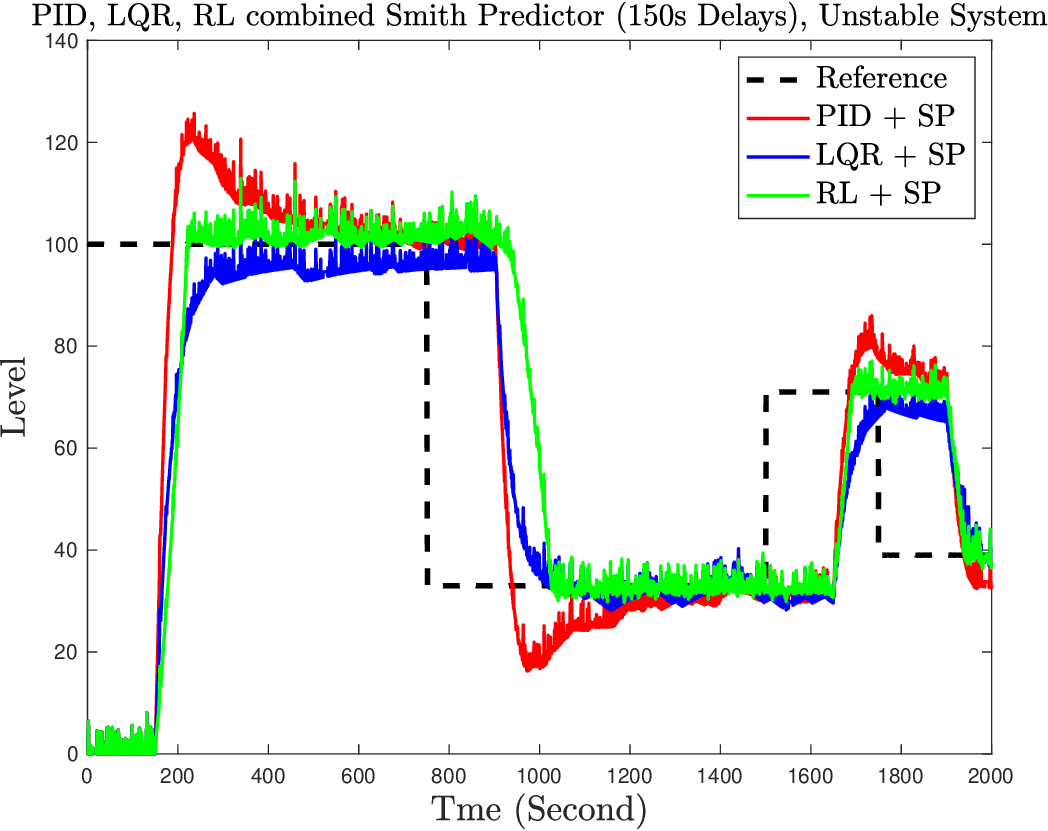}}\\
    \subfloat[
        \label{Ge}\scriptsize
        Stable without SP, $\ell_p=10\,\mathrm{s}$
    ]{\includegraphics[width=.23\linewidth]{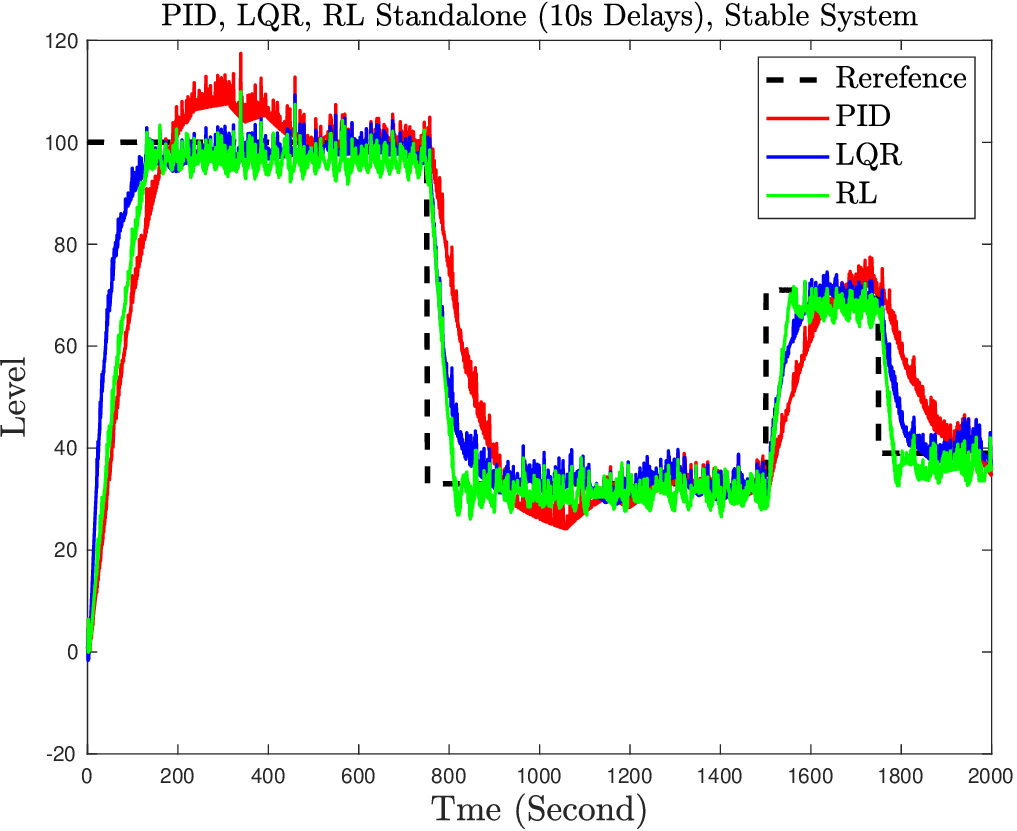}}\;
    \subfloat[
        \label{Gf}\scriptsize
        Stable without SP, $\ell_p=25\,\mathrm{s}$
    ]{\includegraphics[width=.23\linewidth]{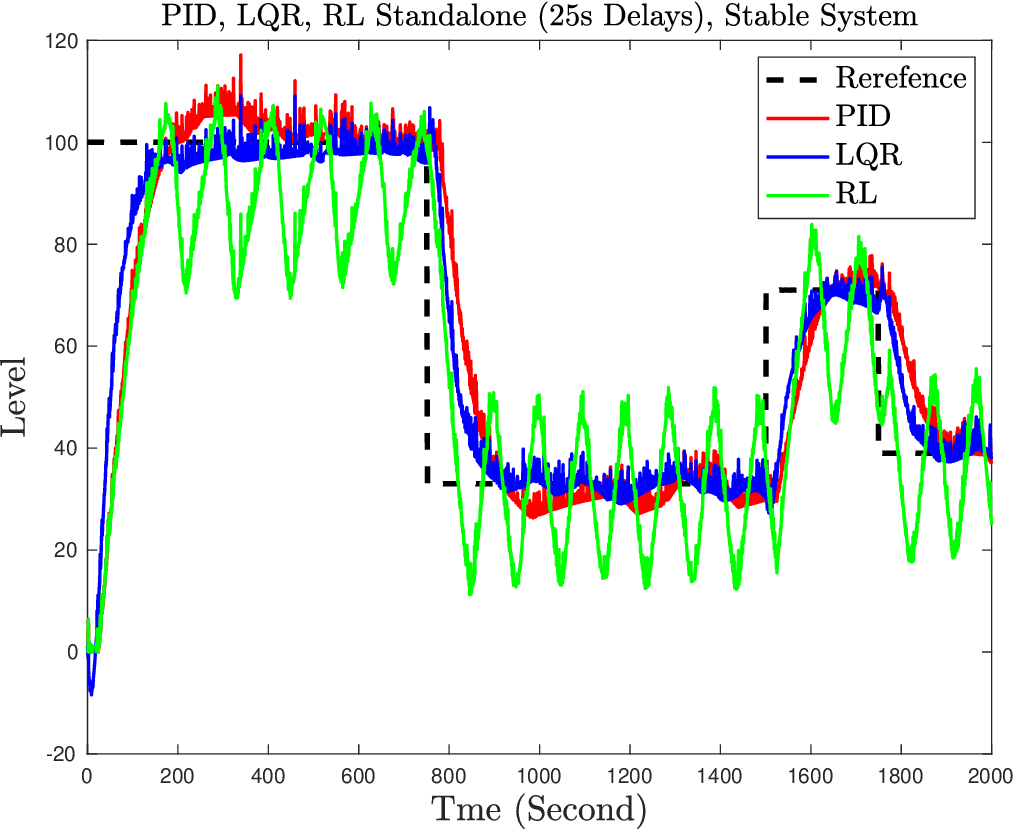}}\;
    \subfloat[
        \label{Gg}\scriptsize
        Stable without SP, $\ell_p=100\,\mathrm{s}$
    ]{\includegraphics[width=.23\linewidth]{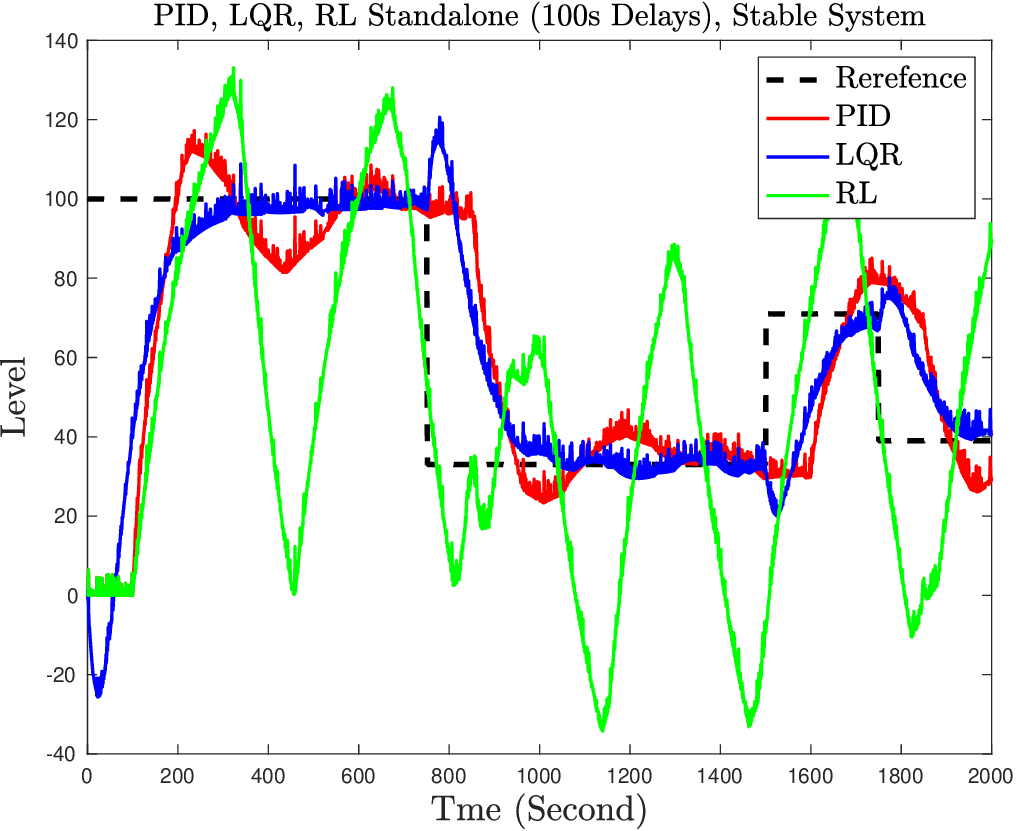}}\;
    \subfloat[
        \label{Gh}\scriptsize
        Stable without SP, $\ell_p=300\,\mathrm{s}$
    ]{\includegraphics[width=.23\linewidth]{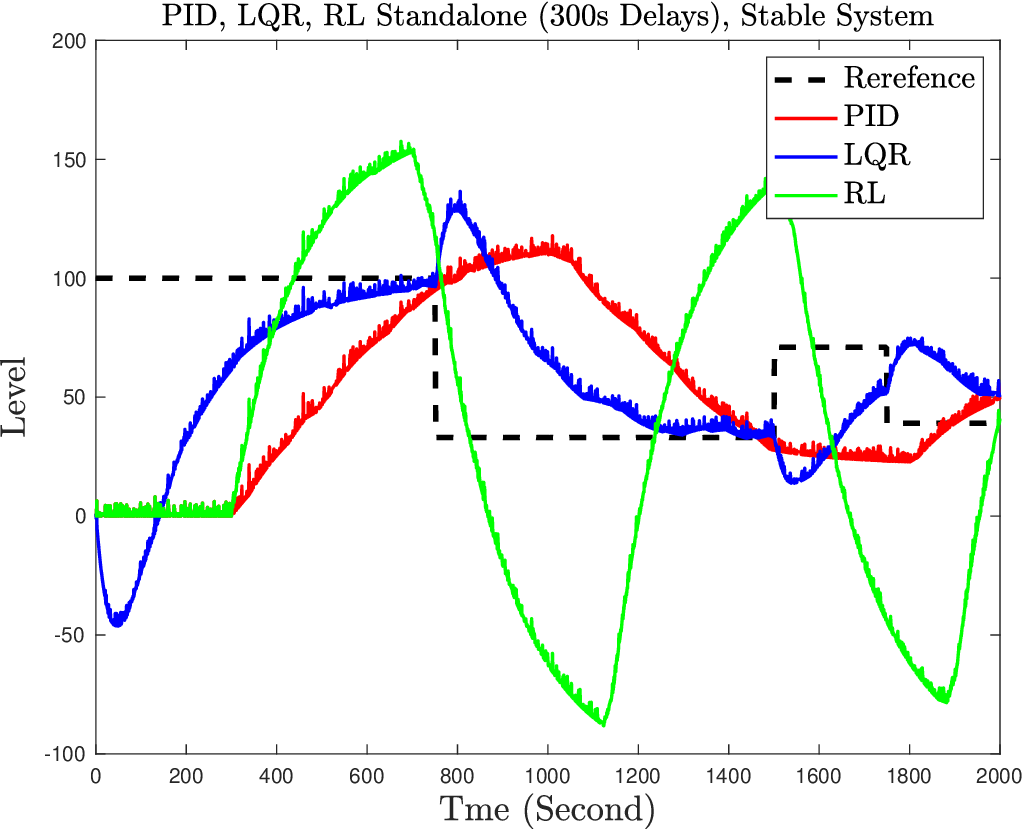}}\\
    \subfloat[
        \label{Gi}\scriptsize
        Unstable without SP, $\ell_p=0\,\mathrm{s}$
    ]{\includegraphics[width=.23\linewidth]{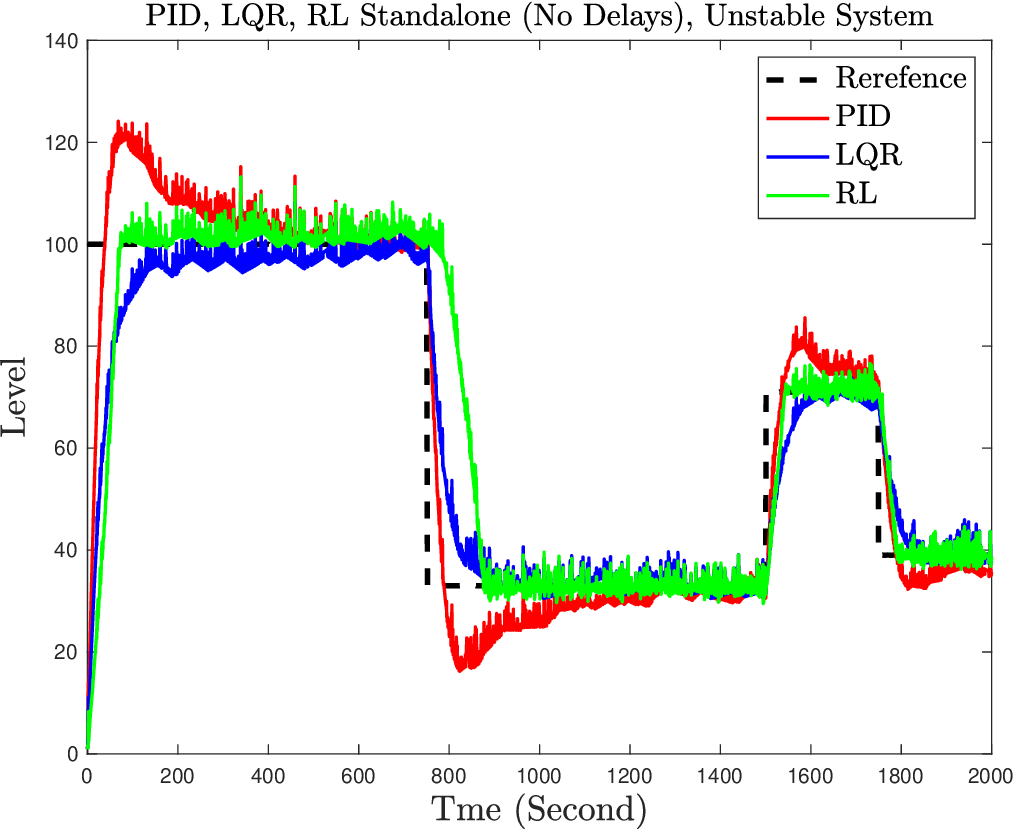}}\;
    \subfloat[
        \label{Gj}\scriptsize
        Unstable without SP, $\ell_p=5\,\mathrm{s}$
    ]{\includegraphics[width=.23\linewidth]{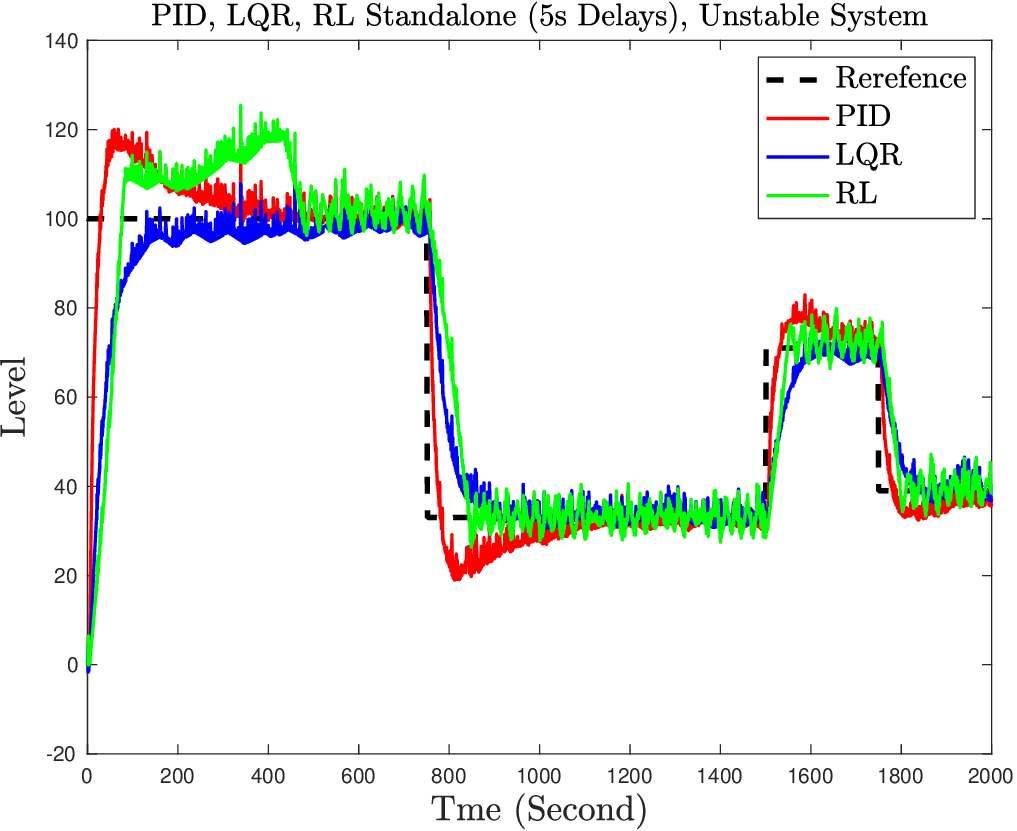}}\;
    \subfloat[
        \label{Gk}\scriptsize
        Unstable without SP, $\ell_p=10\,\mathrm{s}$
    ]{\includegraphics[width=.23\linewidth]{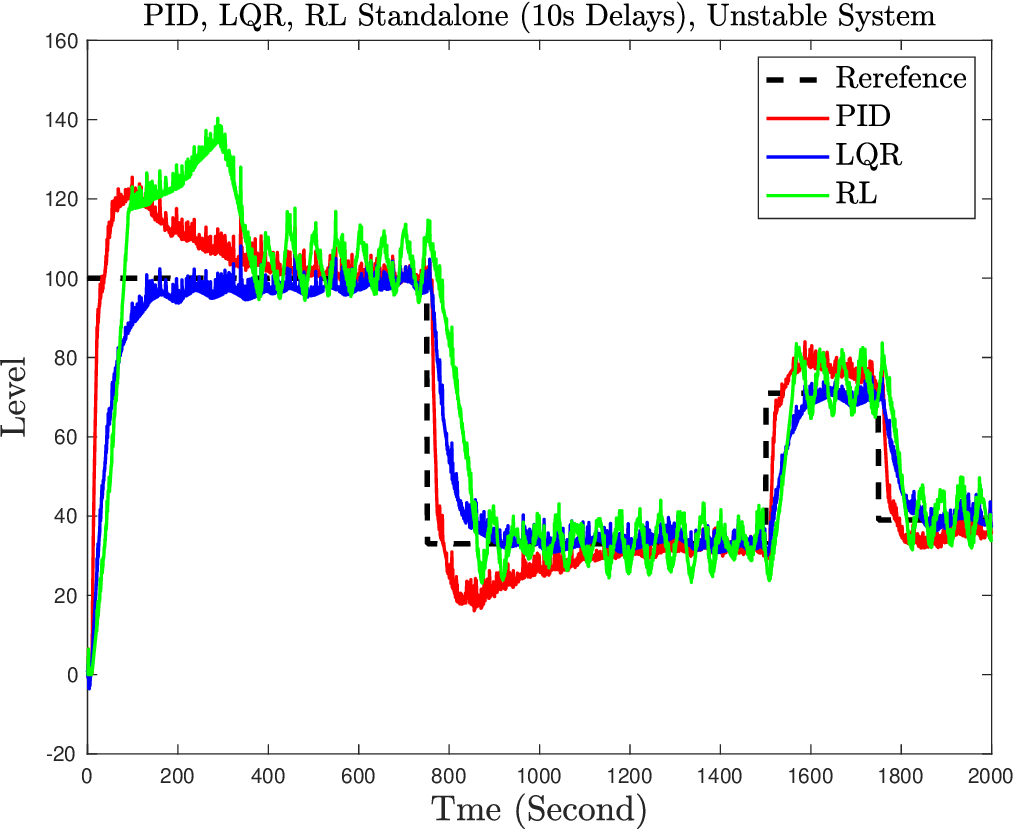}}\;
    \subfloat[
        \label{Gl}\scriptsize
        Unstable without SP, $\ell_p=25\,\mathrm{s}$
    ]{\includegraphics[width=.23\linewidth]{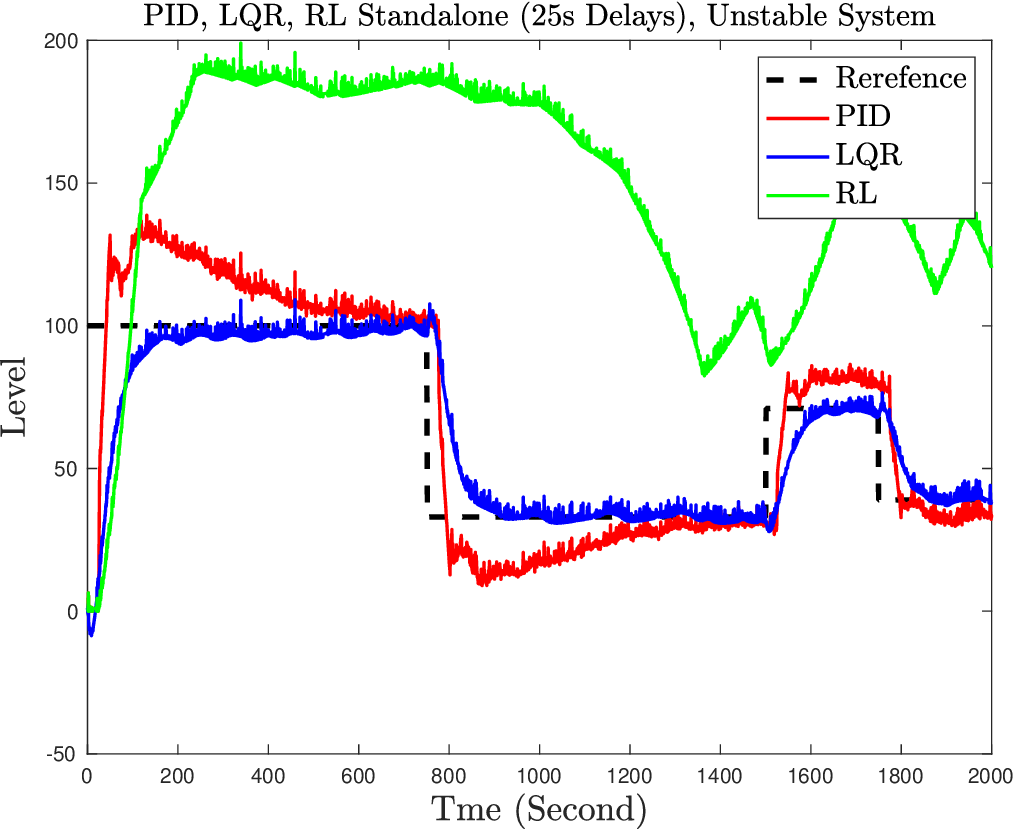}}
    \caption{Tracking responses of PID, LQR, and DDPG
    controllers in the conference experiments.
    (a)--(d) Open-loop unstable plant with Smith-predictor
    compensation; (e)--(h) stable plant without compensation;
    (i)--(l) open-loop unstable plant without compensation.
    The plant delay is indicated in each panel.}
    \label{G1}
\end{figure*}

The delay values considered in the conference study are
\begin{equation}
    \ell_p\in\{5,10,25,50,100,150,300\}\,\mathrm{s},
    \label{eq:simulation-delays}
\end{equation}
with a delay-free case also shown for the unstable plant.
Figure~\ref{G1} reproduces twelve representative responses.
Each panel compares the PID, LQR, and DDPG controllers;
the first row uses Smith-predictor compensation, whereas
the second and third rows show uncompensated feedback.

\subsection{Model separation}

Table~\ref{tab:nugap} reproduces the reported $\nu$-gap
values relative to the delay-free stable benchmark.
For both plant variants, the reported gap is nondecreasing
with delay and is larger for the unstable variant at each
listed delay.

\begin{table}[t!]
    \centering
    \caption{Reported $\nu$-gap values relative to the
    delay-free stable benchmark.}
    \label{tab:nugap}
    \small
    \setlength{\tabcolsep}{4pt}
    \begin{tabular}{c|cc@{\qquad}c|cc}
        \toprule
        $\ell_p$ (s)&Stable&Unstable&
        $\ell_p$ (s)&Stable&Unstable\\
        \midrule
        5  &0.1120&0.4904&100&0.8133&0.9116\\
        10 &0.2069&0.4904&150&0.8913&0.9699\\
        25 &0.4181&0.5251&300&0.9648&0.9999\\
        50 &0.6254&0.7345&   &      &      \\
        \bottomrule
    \end{tabular}
\end{table}

The conference study also reports a gap of $0.4909$
between the stable and unstable variants at equal delay,
and a benchmark robustness quantity of $0.7917$.
These values are retained as reported numerical quantities.
The value $0.7917$ is not used here as a certified margin
for the deployed DDPG policy. Applying
\eqref{eq:nugap-test} requires a verified margin for the
specific linear controller and plant interconnection
being compared.

The gap values describe separation between the linear models.
They do not, by themselves, establish stability of the
nonlinear actor--actuator interconnection. If delay
approximations were used in their computation, their
interpretation is restricted to those approximate models.

\subsection{Tracking responses}

In Figs.~\ref{Ga}--\ref{Gd}, Smith-predictor compensation
substantially reduces the effect of delay on the displayed
responses of the unstable plant. The conference study
reports similar tracking behavior across these delay values.
These finite-duration responses provide numerical evidence
of successful tracking in the tested cases, rather than
a general internal-stability guarantee for unstable plants.
Even under perfect matching, the physical input delay
remains present, as shown by \eqref{eq:perfect-maps}.

For the stable plant without compensation,
Figs.~\ref{Ge}--\ref{Gh} show deterioration as the delay
increases. The conference study reports successful
standalone DDPG operation for delays below approximately
$20\,\mathrm{s}$, with poorer behavior at larger delays.
The tested classical controllers also deteriorate,
with divergent responses reported at
$\ell_p=300\,\mathrm{s}$. These observations concern
the particular controller gains and simulation settings;
they do not define universal delay margins.

Figures~\ref{Gi}--\ref{Gl} show that changing the plant
from stable to open-loop unstable further reduces the
useful deployment range of the fixed actor.
In particular, the reported standalone DDPG response at
$\ell_p=10\,\mathrm{s}$ is poorer than for the stable plant
at the same delay. This observation accompanies the larger
model separation reported in Table~\ref{tab:nugap},
although the gap alone does not predict the nonlinear
policy's tracking performance.

Overall, the retained experiments illustrate the sensitivity
of a fixed learned policy to plant changes and input delay,
and the benefit of Smith-predictor compensation in the
tested configurations. They do not evaluate the delay
identification procedure in Section~\ref{sec:adaptation}
or certify the sufficient conditions of
Section~\ref{sec:robustness}.

\section{Conclusion}
\label{sec:conclusion}

This paper studied Smith-predictor-assisted reinforcement
learning for systems with time delays. The closed-loop
representation showed that perfect matching removes the
delay from the nominal feedback denominator, while the
physical response remains delayed. Plant and delay mismatch
introduce a residual feedback loop, for which small-gain
and $\nu$-gap conditions provide complementary linear
robustness tests. A delay-independent Razumikhin condition
was also established for a fixed Lipschitz policy in a
separate state-feedback configuration.

For delay identification, an output-error gradient was
derived from the nominal-model sensitivity. The projected
gradient flow decreases the fixed-data loss, and local
exponential convergence to the true delay follows under
exact model matching and finite-window excitation.
These identification results do not establish stability
of an online adaptive Smith-predictor loop.

The retained numerical experiments illustrate the benefit
of Smith-predictor compensation for a policy trained on
a stable nominal plant, and the deterioration of standalone
DDPG under increasing delay and plant changes.
The theoretical guarantees depend on their stated
assumptions; neither DDPG training nor nominal delay
compensation alone guarantees closed-loop stability.

\section{Future Work}
\label{sec:future}

First, output-error delay identification could be integrated
with online predictor-based control
\cite{Artstein1982,Wafi-MRAC}. Extending the fixed-data
convergence result requires analyzing the coupled controller
and estimator, with conditions on excitation, estimation
error, and update rates that preserve stability.

Second, the architecture could be extended to multi-agent
systems with heterogeneous delays. Consensus and
event-triggered control
\cite{Dimarogonas2012}, together with
distributed adaptive and nonlinear control
\cite{Wafi-LCSS24,Wafi-ASCC26-DNC}, motivate combining local
predictors with distributed adaptation. The analysis would
need to account for model mismatch, network connectivity,
and delayed neighbor information.

Third, sensor faults could be addressed through fault
diagnosis and cooperative fault-tolerant tracking
\cite{Gao2015FaultDiagnosis,Wafi-ASCC26-FTC}.
This would require distinguishing delay-model errors from
faulty measurements and establishing input-to-state
stability or disturbance-attenuation bounds for the
combined predictor and fault-compensation scheme.

\newpage

\bibliographystyle{IEEEtran}
\bibliography{arXiv}

\end{document}